# Sex-biased expression of microRNAs in *Drosophila melanogaster*

Antonio Marco

School of Biological Sciences, University of Essex, Colchester, United Kingdom

**Address for correspondence:**

School of Biological Sciences

University of Essex

Wivenhoe Park, Colchester CO4 3SQ

United Kingdom

**Email:** amarco.bio@gmail.com

**Telephone:** +44 (0) 120 687 3339

# SUMMARY

Most animals have separate sexes. The differential expression of gene products, in particular that of gene regulators, is underlying sexual dimorphism. Analyses of sex-biased expression have focused mostly in protein coding genes. Several lines of evidence indicate that microRNAs, a class of major gene regulators, are likely to have a significant role in sexual dimorphism. This role has not been systematically explored so far. Here I study the sex-biased expression pattern of microRNAs in the model species *Drosophila melanogaster*. As with protein coding genes, sex biased microRNAs are associated with the reproductive function. Strikingly, contrary to protein-coding genes, male biased microRNAs are enriched in the X chromosome whilst female microRNAs are mostly autosomal. I propose that the chromosomal distribution is a consequence of high rates of *de novo* emergence, and a preference of new microRNAs to be expressed in the testis. I also suggest that demasculinization of the X chromosome may not affect microRNAs. Interestingly, female biased microRNAs are often encoded within protein coding genes that are also expressed in females. MicroRNAs with sex-biased expression do not preferentially target sex-biased gene transcripts. These results strongly suggest that the sex-biased expression of microRNAs is mainly a consequence of high rates of microRNA emergence in the X (male bias) or hitch-hiked expression by host genes (female bias).

# INTRODUCTION

Sexual dimorphism is prevalent in animal species. Sexual phenotypic differences are the consequence of a differential expression of genes between males and females [1]. During the last decade, high-throughput transcript analyses have identified many genes with a sex-biased expression pattern [2–4]. For instance, the *Drosophila* gene *paired* is expressed at a higher level in adult males than in females [5], and it encodes a transcription factor involved in the development of male accessory glands [6]. Indeed, other transcription factors have been identified as sex-biased genes [5,7], indicating that transcriptional gene regulation is tightly linked to sexual dimorphism. Post-transcriptional regulators may also have an impact in sexual dimorphism. MicroRNAs are short endogenous regulatory RNA molecules that are involved in virtually all studied biological processes [8,9]. Recently, differences between male and female microRNA expression profiles have been observed [10–12], suggesting that microRNAs have a role in sexual differentiation.

The study of sex-biased expression of gene products in the model species *Drosophila melanogaster* has produced a number of insightful observations. First, male-biased genes evolve faster that non-biased genes [3,13–15]. Second, the X chromosome is depleted of male based genes and enriched for female biased genes [2,3]. On the other hand, evolutionarily novel genes tend to be X-linked and highly expressed in males [16–23]. These observations suggest a movement of male-biased genes from the X chromosome to the autosomes, a process known as demasculinization of the X [2,20,24]. Therefore, sex-bias expression is an important factor affected the evolutionary fate of protein-coding genes. Likewise, sex-bias expression should have an impact in microRNA evolution. However, this effect may be different to that observed for protein-coding genes as proteins and microRNAs differ in their evolutionary dynamics. For instance, gene duplication is the main mechanism by which novel protein-coding genes emerge, whilst a majority of microRNAs emerge by *de novo* formation within existing transcripts (reviewed in [25,26]). Consequently, novel microRNAs are more likely to be lost than protein-coding genes in a short evolutionary period.

Although microRNAs have been extensively studied in *Drosophila* [27–29], the effect of sex-biased expression in microRNA evolution remains largely unexplored. Here I investigate whether the sexual profile of microRNA expression resembles that of protein-coding genes, and how sex-biased expression affects differently to the evolutionary dynamics of protein-coding genes and microRNAs.

# RESULTS

**Sex-biased expression of *Drosophila* microRNAs**

To characterize which microRNAs have a sex-biased expression pattern in *Drosophila melanogaster*, 13 different small RNA sequencing experiments (including males, females, embryos, ovaries and testes) were crossed-compared (see Methods). Figure 1A shows the correlation among the expression profiles for all experiments, indicating that the female and male pairs of profiles are highly correlated, despite coming from independent experiments. Thus, pairs of male and female profiles were used as biological replicates to calculate differential expression between sexes. A total of 476 mature microRNAs (two sequences per microRNA precursor) were analysed. Of them, 28 and 37 mature microRNAs showed a significant expression bias in males and females respectively (Figure 1B, see Methods). Table 1 includes details of sex-biased microRNAs and their fold change. The expression levels for all analysed mature microRNAs are available in Supplementary Table 1. As only reads mapping to a single microRNA were taken into account, removing reads mapping to multiple sites may influence our analysis. Hence, I compared the expression levels resulting from unique reads and from multiple matching reads. Four microRNA families were affected by multiple matching reads: mir-983, mir-281, mir-276 and mir-2. The first three did not show any differential expression between sexes. However, the fourth one included two microRNAs, mir-2a-1 and mir-13b-2, which are female-biased. To avoid biases due to multiple matches, the mir-2 family was removed from the subsequent analyses.

MicroRNA precursors potentially encode for two mature products (so-called 3 prime and 5 prime products). In agreement with this, many of the sex-biased microRNAs are pairs derived from the same precursor (Table 1). Additionally, microRNAs are frequently clustered in the genome, and these clusters of microRNAs are often transcribed in a single RNA molecule (reviewed in [26]). Indeed, sex-biased microRNAs are frequently clustered, and nearly all of the microRNAs in a cluster show a consistent sex-biased expression (Table 1, Figure 1C). Therefore, the observed bias in mature microRNA production is primarily a consequence of the sex-biased expression of their transcripts.

**Male-biased microRNAs are preferentially located in the X chromosome and expressed in the testis**

Figure 1C shows microRNA transcripts with sex-biased expression and their chromosomal distribution. Contrary to the observation for protein-coding genes, microRNAs expressed in males tend to be located in the X chromosome. In contrast, all female-biased microRNAs are located in autosomes, which is again the opposite observation that has been made for protein coding genes. Figure 1D further explores the relationship between sex-chromosome location and sex-biased expression. The frequency distribution of fold-change in expression for autosomal microRNAs show three peaks, one large of unbiased expression and two smaller of male and female biased expression. However, the distribution of X-linked microRNAs is bimodal (Figure 1D): they are either unbiased or highly- expressed in males. Thus, male-biased microRNAs and the X chromosome are closely associated.

To further understand what it means to be sex-biased expressed, the expression profile of biased microRNAs was explored. Figure 2 plots a hierarchical tree of sex-biased expressed microRNAs and their relative expression levels in testes, ovaries and early embryos. Most male-biased microRNAs are highly expressed in the testes. This indicates that production of microRNAs in males is largely associated with the germ-line and the reproduction function. This is consistent with Figure 1A in which adult samples were poorly correlated with young samples, perhaps because young individuals have not yet developed fully functional gonads.

**Female-biased microRNAs are expressed in ovaries and early embryos**

Figure 2 expression profile shows that female-biased microRNAs falls into three distinct groups. First, a group of female-biased microRNAs are expressed in the somatic stem cells in the ovary, showing that microRNAs are important for the maintenance of stem cells in the ovary, in agreement with previous findings [30]. Second, some female-biased microRNAs are highly expressed in ovaries. This suggests that these microRNAs are important for the formation and maturation of *Drosophila* eggs.

Interestingly, a third group of female-biased microRNAs do not appear to be present in the ovary and they are highly expressed in young embryos (Figure 2). These eggs were originally collected up to one hour after laying [27,31], indicating that these young embryos have not started yet to have zygotic transcription [32]. This suggests that these microRNAs may be maternally deposited by the mother into the unfertilized eggs (oocytes). As a matter of fact, ongoing work in the lab has shown that the microRNAs mir-92a and mir-92b, and the mir-310/mir-311/mir-312/mir-313 cluster, are abundant in *Drosophila* unfertilized eggs (Marco, manuscript in preparation). From these analysis I conclude that both male- and female-biased microRNAs are mostly associated to the reproductive function.

**Intronic female-biased microRNA are associated with host gene expression**

It may be possible that microRNA transcription pattern is associated to the transcription profile of their neighbouring protein-coding genes. In general, as shown in Table 2, both expression patterns were not significantly associated (11 out of 19 microRNA transcripts have the same expression bias as their closest neighbouring gene; p=0.32, binomial test). In particular, there are 8 microRNA transcripts with male-biased expression, and only 3 of their respectively closest genes show a similar bias. In the case of female-biased microRNA transcripts, 8 out of 11 have their closest gene with a female-biased expression pattern. A closer inspection to the data reveals that this bias is produced mostly by microRNAs hosted within protein coding genes (overlapping transcripts). Indeed, all 6 genes hosting microRNAs with female-biased expression are themselves higher expressed in females than in males (p=0.016). This shows that female biased expression of microRNAs is highly associated to the production of microRNAs from introns of sex-biased expressed protein-coding genes.

**Evolutionary origin of sex-biased microRNAs**

There are two possible ways a gene may become sex-biased. First, a gene can acquire sex-biased expression. Second, a new gene appears (either *de novo* or by the duplication of an existing gene) having from the very beginning a sex-biased expression. Figure 4 shows the evolutionary origin of sex-biased microRNAs. Most male-biased microRNAs emerged within the *Drosophila* lineage, with only two exceptions: mir-993 and mir-283/304/12. These are indeed the least biased of all of the microRNAs. Thus, microRNAs with a strong male bias are evolutionarily young. In contrast, the evolutionary origin of female-biased microRNA families is diverse, and there are both old and young microRNAs. Among the old microRNAs we have the mir-92, mir-184 and mir-9 families, which are conserved even in chordates. Interestingly, there are no *Drosophila melanogaster* specific microRNAs with a clear female-bias expression (contrary to the case of male-biased microRNAs). There are, however, two female-biased microRNAs which appeared in the *Drosophila* genus lineage: mir-314 and the mir-310~mir-313 cluster.

**Targets of sex-biased microRNAs**

Do sex-biased microRNAs also target sex-biased expressed gene transcripts? To explore this question three different target prediction algorithms were used: TargetScan, miRanda and DIANA-microT (see Methods section). MicroRNAs were binned by their bias level, and the expression bias of their targets were plotted in Figure 3. These boxplots show that there is no tendency of sex-biased microRNAs to target sex-biased transcripts, at least not as a global pattern. I further explored the targets of seven melanogaster-subgoup-specific male-biased microRNAs. For two of them, 2 out of the 3 prediction algorithms detected a significant association with sex-biased transcripts: mir-985 have a tendency to target female-biased genes, whilst mir-997 significantly target male-biased genes (Supplementary Table 2). The other associations were not significant and/or supported by only one prediction algorithm. Last, I investigated whether recently-emerged male-biased microRNAs also target evolutionarily young genes. I calculated the ratio between *Drosophila*-specific and conserved targeted genes for the targets predicted for the three above-mentioned algorithms. Four out of seven studied microRNAs showed a tendency to target more conserved genes than expected by chance for at least two algorithms (Supplementary Table 3), among them, mir-985. The results here described rely heavily based on target prediction algorithms and, therefore, should be taken with caution. However, they suggest that newly emerged microRNAs can potentially target conserved genes, altering regulatory relationships that have been conserved throughout evolution.

## DISCUSSION

In this study I have shown that sex-biased microRNAs are mainly associated with the reproductive function: male microRNAs are expressed in testes and female microRNAs are abundant in ovaries and oocytes. However, their evolutionary origin is different. Male biased microRNAs tend to be evolutionarily young (dipteran/*Drosophila* specific; Figure 4) and they often emerge in the X chromosome. Contrary to microRNAs, male biased protein-coding genes appear to be generally under-represented in the X chromosome in flies [2,3], and a movement of male genes out of the X, or demasculinization of the sex chromosome, has been suggested [2,20]. However, novel genes tend to be X-linked and male expressed and older genes may have moved outside the X chromosome [2,3,17,21]. An enrichment in the X chromosome for microRNAs with male-biased expression has also been reported in mammals [12,33–35].

A careful dissection of the evolutionary origin of male-biased genes in *Drosophila* demonstrated that *de novo* originated genes tend to be X-linked and male-biased, and that there may also be an ongoing demasculinization process in the X chromosome [23]. The study also suggested that this demasculinization may also be happening in microRNAs. Their analysis showed that there is about a 12 fold enrichment of evolutionarily young microRNAs in the X with respect to autosomes. For conserved microRNAs the enrichment is less than 2 fold. However, when taken into account that multiple microRNAs may come from the same transcript (Figure 1C) the figures are different: 3.5 and 1.8 fold enrichment for young and conserved microRNAs respectively. These differences are small, and evidence for demasculinization in microRNAs is not supported.

There is an ongoing debate in the scientific literature about sex chromosome demasculinization. Although demasculinization has been generally considered one of the prominent features of *Drosophila* X chromosome evolution, recent work shows that the observed paucity of male-biased genes in the X chromosome may be artifactual [36–38]. Indeed, several groups suggest that demasculinization does not happen in *Drosophila* and propose that there is no global meiotic sex chromosome inactivation, or MSCI [39,40]. The movement of male biased genes out of the X is often explained as a response to MSCI. This discussion is not settled and evidence for both demasculinization and MSCI is still reported [23,41–43]. Interestingly, most X-linked microRNAs escape MSCI [44]. These observations imply that X-chromosome demasculinization caused by MSCI might not happen during microRNA evolution. Even if there is an ongoing demasculinization process affecting protein-coding genes, microRNAs seem not to be affected.

Female microRNAs are generally older than male biased microRNAs, and they are frequently encoded within other female expressed genes. For instance, mir-995 is highly expressed in females (Figure 1) and it is associated to oocytes (Figure 2). This microRNA is encoded within the first intron of *cdc2c*, a gene involved cell proliferation during development [45]. Hence, the presence of mir-995 in oocytes may be a by-product of the host gene expression. Also, mir-995 can be identified in the same intron of the orthologous *cdc2c* gene in other insects [46], showing a deep conservation of the microRNA/host gene association. Interestingly, mir-92a is encoded within a gene (*jigr1*) whose product is maternally deposited in the oocyte [47], and the microRNA is highly expressed in oocytes (Figure 2) and detected in unfertilized eggs (Marco, manuscript in preparation). The presence of mir-92a in the developing egg may be a by-product of being intronic of a sex-biased expressed gene. As a matter of fact, mir-92a is associated to leg morphological differences between *Drosophila* species [48], a role (in principle) unrelated with any function in the early developing egg.

Among the microRNAs with a female-biased expression pattern there are microRNAs associated to the gametic function. Recently, mir-989 has been discovered to be involved in cell migration in the ovary [49]. Indeed, the 3' arm of mir-989 is highly expressed in ovaries (Figure 2). The analysis of female mutants also reveals that mir-9c (present in ovaries; Figure 2) is somewhat involved in the control of the number of germ cells [50]. Predictably, other female-biased microRNAs here reported, such as mir-994/318, could have a role in gametic function. Strikingly, the mir-310/311312/313, which is female expressed (and probably maternally deposited in the egg), is involved in the development of male gonads [51]. This emphasizes that genes with sex-biased expression can also have other functions, even in the opposite sex.

We recently characterized sex-biased microRNAs in the parasitic *Schistosoma mansoni* and reported that one of the microRNA clusters (mir-71/mir-2) have two copies, one in the sexual chromosome with no detectable bias and another copy in an autosome with sex-bias expression. The duplication of the cluster happened more or less at the same time sexual dimorphism appeared in this genus (*Schistosoma*). We suggested that this may be a case of escaping sex conflict, in which genes involved in sex dimorphism tend to be out of the X [10,52]. However, this is likely to be an exception to the rule in microRNAs, as their evolutionary dynamics is primarily dominated by high levels of emergence and a low probability of non-tandem duplication.

In summary, I conclude that sex-bias expression of microRNAs is a consequence of a high rate of microRNA *de novo* emergence. Novel microRNAs tend to appear in the X chromosome and to be expressed in the testes. Conversely, male biased microRNAs are evolutionarily young and show a high rate of loss as well. On the other hand, many female biased microRNA emerged within the intron of female-biased host genes. They are generally conserved suggesting that the female gametic function may be more constrained, and purifying selection could eliminate emerging microRNAs impairing ovary/oocyte development. This scenario suggests that positive/adaptive selection may have nothing but a little contribution on determining the sex-biased expression of microRNAs.

## METHODS

*Drosophila melanogaster* microRNA sequences are from miRBase version 19 [53]. Expression datasets were downloaded from Gene Expression Omnibus (GEO) at http://www.ncbi.nlm.nih.gov/geo/, with accession numbers: GSM286602 and GSM399107 [adult males]; GSM286603 and GSM399106 [adult females]; GSM280082 [ovaries]; GSM909277 and GSM909278 [testes]; GSM385822 and GSM385744 [ovary somatic sheet]; GSM180330 and GSM286613 [early embryos]; GSM609223 and GSM609224 [young males and females] [31,54–58]. Reads from these experiments were mapped to *Drosophila melanogaster* microRNA hairpins with Bowtie 0.12.7 [59], allowing no mismatches nor multiple matches. Differential expression of microRNAs was estimated with edgeR [60]. In short, read counts were first normalized with the TMM method [61]. Then, the variation within samples is estimated by fitting the expression pattern to a negative binomial distribution. Sex-biased microRNAs are detected by an exact test controlling the false discovery rate [62]. Additionally, the analysis was repeated with a more general method implemented in DESeq [63], but the results remained overall the same. Expression data for *Drosophila* genes were obtained from modENCODE, available at flybase.org [57,64]. All statistical analyses and Figures 1 and 2 were done with R [65].

Evolutionary age of microRNAs and microRNA families were estimated as previously described [66]. In brief, I compiled microRNA sequences with detectable similarity to *D. melanogaster* microRNAs with BLAST [67], using a sensitive set of parameters to detect homologous microRNAs (-w 4, -q -3, -r +2). I also included additional sequences described elsewhere, to ensure that all known microRNA families with a common evolutionary origin are taken into account [53,66,68–70]. MicroRNA hairpins were aligned with ClustalX 2.0 [71], manually refining the alignments with Ralee [72] and phylogenetic trees were reconstructed using the neighbor-joining and maximum likelihood routines with default parameter as implemented in MEGA 5 [73]. MicroRNA age estimates were also compared to those obtained by Mohammed et al [74] using a different approach: they analysed whole genome alignments of 12 *Drosophila* genomes [75]. Age estimations were fully congruent between both datasets.

MicroRNA targets were retrieved from our previous study [76]. In short, 3'UTRs were downloaded from FlyBase (genome version BDGP 5.25), and the microRNA targets were predicted with three different programs based on different algorithmic approaches: TargetScan [77], DIANA-microT [78], and miRanda [79], with default parameters. The evolutionary conservation of targeted genes was inferred from the gene family tree available at TreeFam 9 [80].

**Table 1.** MicroRNA mature sequences with sex-biased expression

| Female-biased | | | Male-biased | |
|---|---|---|---|---|
| mir-989-5p (9.8) | mir-995-5p (3.4) | mir-13b-2-5p (1.7) | mir-985-3p (8.8) | mir-978-3p (4.7) |
| mir-994-5p (8.7) | mir-313-3p (3.2) | mir-314-5p (1.7) | mir-976-3p (7.7) | mir-959-5p (4.5) |
| mir-989-3p (8.7) | mir-310-3p (3.1) | mir-306-5p (1.6) | mir-991-3p (7.2) | mir-960-3p (4.2) |
| mir-994-3p (8.2) | mir-279-5p (3.0) | mir-79-3p (1.4) | mir-977-5p (7.2) | mir-961-5p (4.1) |
| mir-318-3p (8.2) | mir-995-3p (2.9) | mir-996-5p (1.3) | mir-978-5p (6.8) | mir-303-5p (4.1) |
| mir-310-5p (7.8) | mir-79-5p (2.9) | mir-308-5p (1.1) | mir-4966-5p (6.6) | mir-984-5p (3.4) |
| mir-318-5p (6.7) | mir-9c-3p (2.5) | mir-996-3p (1.1) | mir-973-5p (5.9) | mir-959-3p (3.2) |
| mir-92a-3p (5.7) | mir-9c-5p (2.5) | mir-184-3p (1.0) | mir-975-5p (5.8) | mir-963-5p (3.0) |
| mir-313-5p (5.1) | mir-9b-5p (2.4) | mir-279-3p (1.0) | mir-982-5p (5.7) | mir-303-3p (3.0) |
| mir-92b-3p (5.0) | mir-312-5p (2.3) | | mir-997-5p (5.5) | mir-964-5p (2.8) |
| mir-312-3p (4.3) | mir-9b-3p (2.2) | | mir-972-3p (5.3) | mir-960-5p (2.5) |
| mir-311-3p (3.9) | mir-92a-5p (2.2) | | mir-961-3p (5.1) | mir-2a-1-3p (1.5) |
| mir-92b-5p (3.6) | mir-998-3p (1.9) | | mir-977-3p (5.0) | mir-993-3p (1.3) |
| mir-311-5p (3.4) | mir-2a-1-5p (1.8) | | mir-992-3p (4.8) | mir-12-5p (1.1) |

**Table 2.** Sex-biased microRNA transcripts and their closest neighbouring genes

| MicroRNA/cluster | fold change[1] | distance to closest gene[2] | closest gene | fold change[1] |
| --- | --- | --- | --- | --- |
| mir-972~mir-979 | -5.2 | overlapping | Grip84 | 2.3 |
| mir-982~mir-984 | -3.5 | overlapping | CG3626 | 0.4 |
| mir-959~mir-964 | -2.8 | overlapping | CG31646 | -2.6 |
| mir-283~mir-12 | -1.1 | overlapping | Gmap | -0.3 |
| mir-985 | -6.7 | 19344 | disco | -1.5 |
| mir-997 | -5.3 | 737 | D1 | 1.2 |
| mir-2498~mir-992 | -4.7 | 817 | CG32532 | 0.0 |
| mir-993 | -1.2 | 11072 | Ama | 2.0 |
| mir-92a~mir-92b | 4.6 | overlapping | jigr1 | 2.6 |
| mir-995 | 3.0 | overlapping | cdc2c | 3.9 |
| mir-9c~mir-9b | 2.2 | overlapping | grp | 3.1 |
| mir-184 | 1.1 | overlapping | CG44206 | 0.0 |
| mir-308 | 1.0 | overlapping | RpS23 | 1.2 |
| mir-998~mir-11 | 0.4 | overlapping | E2f | 1.4 |
| mir-989 | 8.6 | 2739 | Rcd1 | 0.9 |
| mir-994~mir-318 | 8.3 | 249 | Irp-1B | -0.3 |
| mir-310~mir-313 | 4.1 | 1624 | gsm | -0.7 |
| mir-279~mir-996 | 1.1 | 2892 | Ef1gamma | 1.4 |
| mir-314 | 0.2 | 182 | Tim13 | -7.5 |

[1]logarithm of fold change between males and females expression levels
[2]in nucleotides

**FIGURE LEGENDS**

**Figure 1. Sex-biased microRNAs in *Drosophila melanogaster*.** (A) Heatmap of cross-correlations of all expression datasets analysed. Different experiments are hierarchically clustered. (B) Smear plot of mature microRNA sequences. Gray lines indicate a 2-fold difference in expression levels between males and females. Red dots are microRNAs with a statistically significant differential expression. (C) MicroRNA transcripts with sex-biased expression, average fold-change of their products and their chromosomal location. (D) Frequency plot of sex-biases in expression levels for autosomes and the X chromosome.

**Figure 2. Expression profile of sex-biased microRNAs.** Hierarchical clustering and heatmap of microRNAs with sex-biased expression. Z-scores were scaled across rows. Green colour indicates an overexpression in a given tissue/sample with respect to the other samples (columns).

**Figure 3. Expression bias of microRNAs and their targets.** Box-plots of expression bias of gene transcripts targeted by microRNAs with no (0), moderate (-5/5) and large (-10/10) sex biased expression. Targets are shown for three different target prediction algorithms: TargetScan, DIANA-microT and miRanda.

**Figure 4. Evolutionary origin of sex-biased microRNA transcripts.** Phylogenetic tree of *Drosophila melanogaster* and other animal groups. MicroRNAs emerging at a given lineage were shown over the relevant branches. For microRNA clusters, only the first microRNA is shown in the figure over the branch at which the oldest microRNA emerged. Red microRNAs are female-biased and blue are male-biased.

# Figure 1

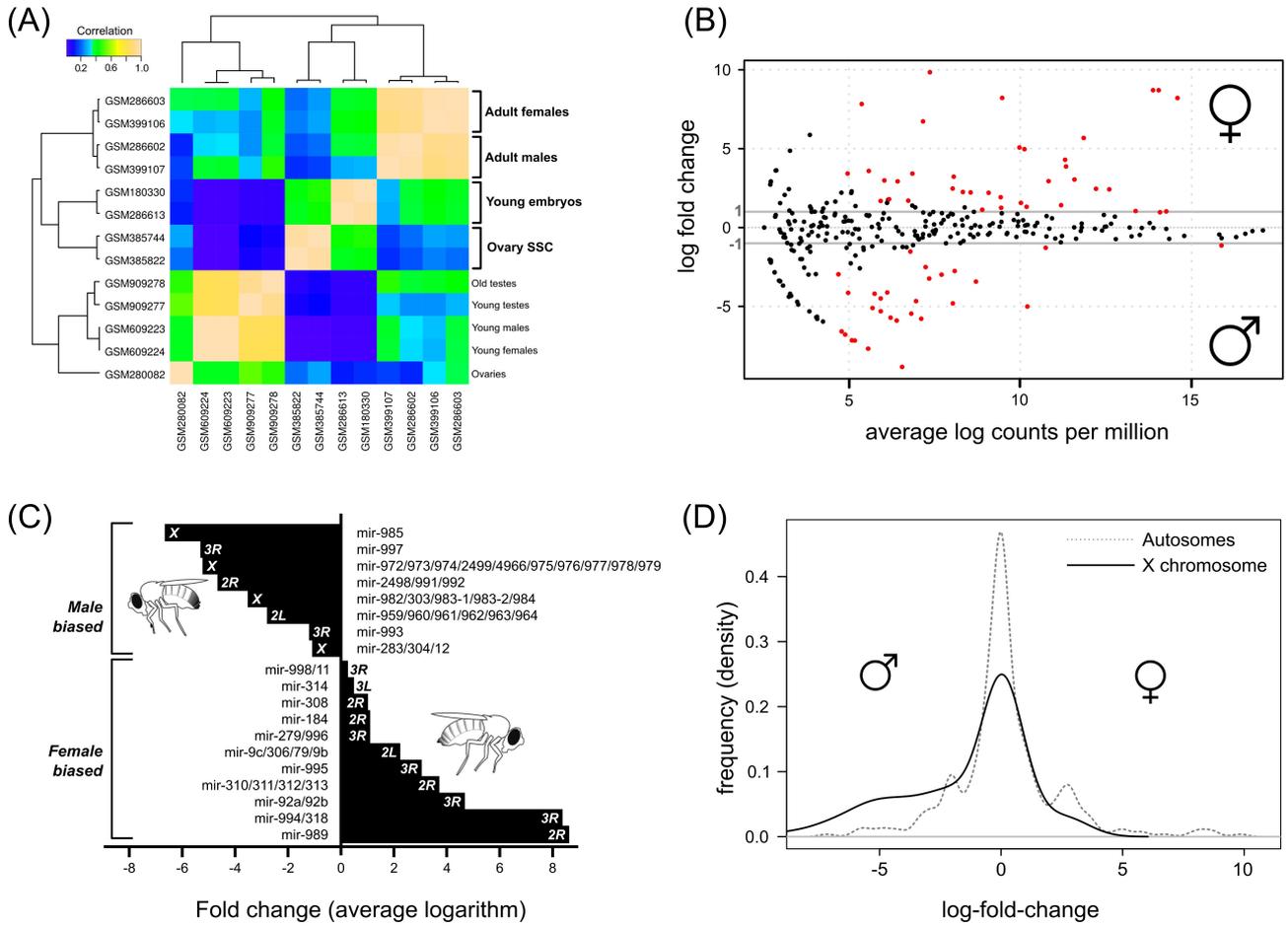

**Figure 2**

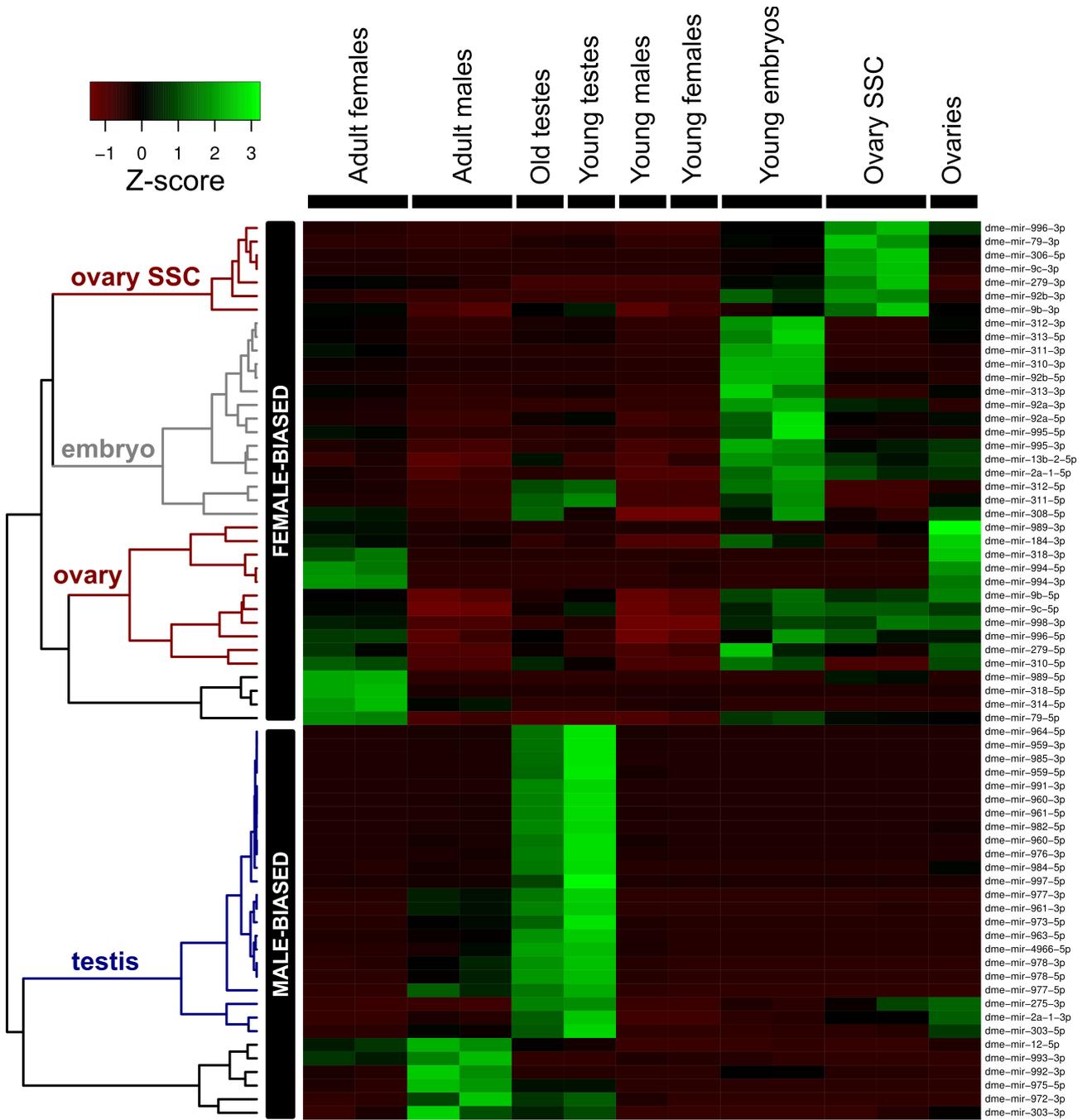

**Figure 3**

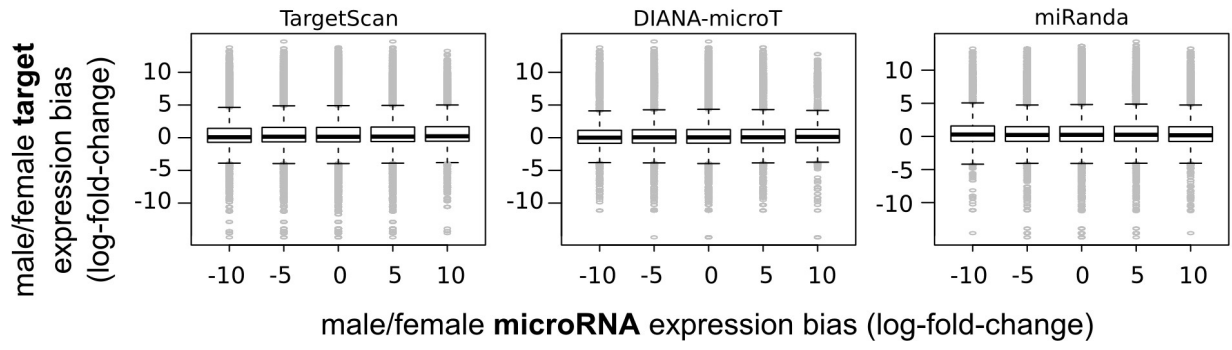

**Figure 4**

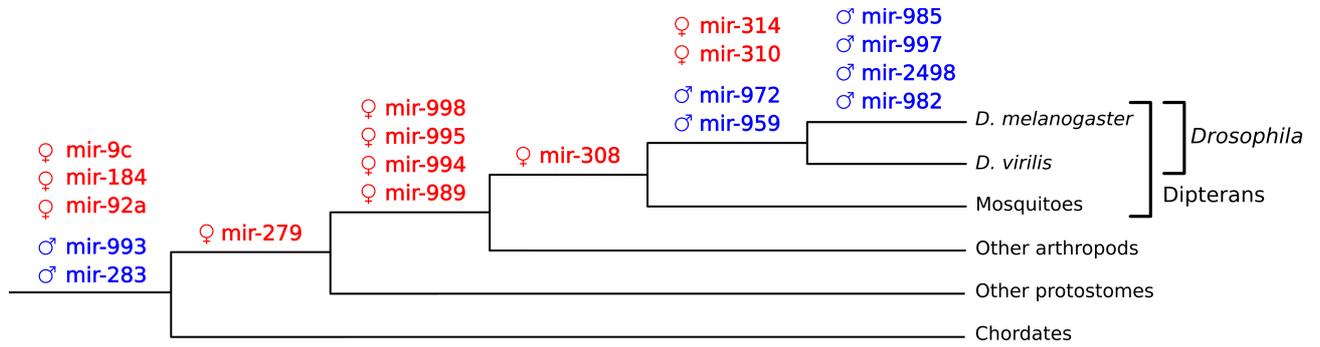